\documentclass[fleqn,usenatbib]{mnras}

\usepackage{newtxtext,newtxmath}

\usepackage[T1]{fontenc}

\DeclareRobustCommand{\VAN}[3]{#2}
\let\VANthebibliography\thebibliography
\def\thebibliography{\DeclareRobustCommand{\VAN}[3]{##3}\VANthebibliography}

\usepackage{mathtools}	

\usepackage{graphicx}	
\usepackage{amsmath}	
\usepackage{amssymb}	
\usepackage[utf8]{inputenc}
\usepackage[english]{babel}
\usepackage{booktabs}
\usepackage{siunitx}
\usepackage[normalem]{ulem}

\newcommand{\orcid}[1]{\href{https://orcid.org/#1}{\includegraphics[width=8pt]{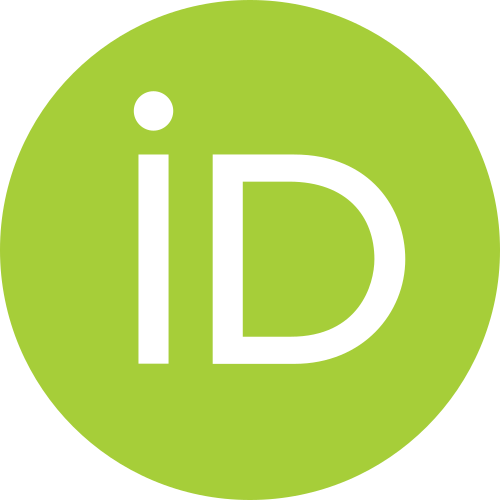}}}
\newcommand{\QR}{$Q_{\textnormal{R}}$ }
\newcommand{\QRD}{$Q_{\textnormal{RD}}^{*}$ }

\graphicspath{{./}{plots/}}

\DeclareSIUnit\Mearth{M_\oplus}
\DeclareSIUnit\Rearth{R_\oplus}
\DeclareSIUnit\erg{erg}

\title[Forming Iron-rich Planets with GI]{Forming Iron-rich Planets with Giant Impacts}

\author[C. Reinhardt et al.]{
Christian Reinhardt\orcid{0000-0002-4535-3956}$^{1}$\thanks{E-mail:christian.reinhardt@ics.uzh.ch},
Thomas Meier\orcid{0000-0001-9682-8563}$^{1}$,
Joachim G. Stadel\orcid{0000-0001-7565-8622}$^{1}$,
Jon F. Otegi$^{1,2}$,
Ravit Helled\orcid{0000-0001-5555-2652}$^{1}$
\\
$^{1}$Institute for Computational Science, University of Zurich, Winterthurerstrasse 190, 8057 Zurich, Switzerland\\
$^{2}$Observatoire Astronomique de l’Universit\'e de Gen\`eve, 51 Ch. des Maillettes, – Sauverny – 1290 Versoix, Switzerland\\
}

\date{Accepted XXX. Received YYY; in original form ZZZ}

\pubyear{2022}

\begin{document}
\label{firstpage}
\pagerange{\pageref{firstpage}--\pageref{lastpage}}
\maketitle

\begin{abstract}
We investigate mantle stripping giant impacts (GI) between super-Earths with masses between \SI{1}{\Mearth} and \SI{20}{\Mearth}. We infer new scaling laws for the mass of the largest fragment and its iron mass fraction, as well as updated fitting coefficients for the critical specific impact energy for catastrophic disruption, $Q_{\textnormal{RD}}^{*}$. With these scaling laws, we derive equations that relate the impact conditions, i.e., target mass, impact velocity and impactor-to-target mass ratio, to the mass and iron mass fraction of the largest fragment. This allows one to predict collision outcomes without performing a large suite of simulations. Using these equations we present the maximum and minimum planetary iron mass fraction as a result of collisional stripping of its mantle for a given range of impact conditions. We also infer the radius for a given mass and composition using interior structure models and compare our results to observations of metal-rich exoplanets. We find good agreement between the data and the simulated planets suggesting that GI could have played a key role in their formation. Furthermore, using our scaling laws we can further constrain the impact conditions that favour their masses and compositions. Finally, we present a flexible and easy-to-use tool that allows one to predict mass and composition of a planet after a GI for an arbitrary range of impact conditions which in turn allows to assess the role of GI in observed planetary systems.
\end{abstract}



\section{Introduction}
\label{sec:introduction}
Since the discovery of the first exoplanet around a main-sequence star at the end of the last century \citep{Mayor1995}, over 5000\footnote{NASA Exoplanet Archive: \url{https://exoplanetarchive.ipac.caltech.edu/}} planets orbiting other stars than the Sun have been discovered. Their compositions are very diverse and range from bodies that are dominantly made of refractory elements to ones that are mostly composed of hydrogen and helium. Planets that have a composition similar to Earth, i.e., consist mostly of iron and rock, seem to be very common (e.g., \citealt{Petigura2013, Bryson2020}). Among those rocky planets, several have very high bulk densities implying that they are enriched in iron compared to the chondritic value and could have a composition similar to Mercury. Some examples for such planets are K2-38b \citep{Toledo-Padron2020}, Kepler-107c \citep{Bonomo2019}, K2-229b \citep{Santerne2018}, L 168-9b \citep{Astudillo-Defru2020}, K2-106b \citep{Guenther2017}, Kepler-80d \citep{Hadden2017}, K2-291b \citep{Kosiarek2019}, Kepler-406b \citep{Marcy2014} and the recently discovered GJ 367b \citep{Lam2021} (see Table~\ref{tab_ExoplanetProperties} for masses, radii and compositions). Their masses range from \SI{0.55}{\Mearth} to \SI{9.4}{\Mearth} so they are at least one order of magnitude more massive than Mercury and therefore are often called super-Mercuries.

The origin of super-Mercuries is still unknown and several formation mechanisms have been proposed. In a recent study \citet{Adibekyan2021} find a link between the compositions of rocky planets and their host stars. In their sample of exoplanets super-Mercuries are found to be more common around stars that are enriched in iron. However, this is not the case for all iron-rich planets. For example, the extreme composition of Kepler-107c cannot be explained by the stellar abundance of rock-building elements \citep{Schulze2021}. Furthermore, obtaining the high metal abundance of super-Mercuries would require an unrealistically large iron enrichment of the host star. The proto-planetary disc from which super-Mercuries are formed could be locally enriched in iron, e.g., due to photophoretic forces \citep{Wurm2013} or different condensation temperatures of silicates and iron \citep{Lewis1972, Aguichine2020}, but this usually requires very specific disc conditions. It is unclear whether such conditions were met in the observed planetary systems. Another possibility is that part of the mantle is lost due to the photo-evaporation of silicates \citep{Santerne2018}. However, this mechanism is limited to planets orbiting very close to the host star due to the extreme temperatures required to evaporate rock. In addition, photo-evaporation is unlikely to remove more than a few percent of the mantle for planets with masses larger than the Earth \citep{Ito2021}. It is therefore difficult to explain a Mercury-like composition by photo-evaporation alone. 

Finally, giant impacts (GI) could strip a large fraction of the mantle analogous to the scenario that was proposed for Mercury in the Solar system (e.g., \citealt{Benz2007, Asphaug2014a, Chau2018, Clement2021}), leaving behind super-Mercuries.
All of the super-Mercuries listed above orbit very close (between \SI{0.01}{\astronomicalunit} and \SI{0.06}{\astronomicalunit}) to their host star and therefore have orbital velocities between \SI{130}{\kilo\meter\per\second} and \SI{260}{\kilo\meter\per\second}. As a consequence, high velocity impacts during the formation and early evolution of these planetary systems are possible. This in turn promotes disruptive collisions that can strip a large fraction of the mantle, increasing the planet's iron mass fraction. Indeed, in the case of Kepler-107c \citet{Bonomo2019} demonstrated that GI is the likely origin of its iron-rich composition using 3-D hydro simulations to model the collision.

In the general context of super-Earths \citet{Marcus2009} (hereafter M09) investigate mantle stripping GI using 3-D hydro simulations and derive scaling laws for the mass of the largest fragment as well as its iron mass fraction, $Z_{\textnormal{Fe}}$. The simulations focus on head-on impacts since these impacts are the most erosive and grazing collisions occurring above the mutual escape velocity tend to be hit-and-run collisions \citep{Leinhardt2012, Timpe2020}. Using these scaling laws \citet{Marcus2010a} (hereafter M10) then derive equations that relate the impact conditions, e.g., the impact velocity, impactor-to-target mass ratio and total colliding mass, to the fragment mass and iron mass fraction. This allows one to predict the fragment mass and composition for given impact conditions without performing impact simulations. Furthermore, they calculate each fragment's radius from its mass and composition using the interior models for super-Earths of \citet{Valencia2007}. From these results the minimum radius of a planet due to collisional stripping of its mantle is inferred assuming an upper limit of \SI{80}{\kilo\meter\per\second} for the impact velocity. This result is often used for interpreting the role of GI when explaining the composition of observed exoplanets. Additionally, M10 also predict a maximum mass of \SI{5}{\Mearth} for super-Mercuries based on the scaling laws and the assumption that the initial pre-impact mass of a super-Earth does not exceed \SI{10}{\Mearth}. However, several substantially more massive iron-rich planets like Kepler-107c or K2-106b have been observed. Furthermore, all of the iron-rich exoplanets presented in Table~\ref{tab_ExoplanetProperties} orbit very close to their host star so their orbital velocities exceed \SI{80}{\kilo\meter\per\second} allowing much larger impact velocities than assumed in M10. Finally, M10 make several mistakes when deriving their equations (see Appendix~\ref{appendix:discussion_results_m10} for details). This in turn results in errors in the minimum radius of exoplanets that are comparable to the accuracy of current observations. We therefore reexamine such collisions considering higher pre-impact masses and impact velocities.

In this work we use 3-D Smoothed Particle Hydrodynamics (SPH) simulations to investigate mantle stripping in GI between super-Earths with masses of \SI{1}{\Mearth} to \SI{19}{\Mearth}. We then derive scaling laws for the mass of the largest fragment and its iron mass fraction that are applicable over a wide range of impact conditions, particularly higher masses and impact energies, as well as updated fitting coefficients for the critical specific impact energy for catastrophic disruption, \QRD. Furthermore, we combine these scaling laws to obtain a general relationship between the impact conditions, i.e., target mass, impactor-to-target mass ratio and impact velocity, and the post-impact fragment mass and composition following M10. For each fragment mass and composition we also determine the radius, making the connection to observational data. We then present a mass-radius (M-R) diagram where we compare the results of our simulations to the iron-rich exoplanets listed in Table~\ref{tab_ExoplanetProperties} and show that their compositions are consistent with mantle stripping GI. Finally, we propose a general criterion to constrain the minimum and maximum iron mass fraction possible for exoplanets via the process of collisional stripping based on the expected impact conditions and the total colliding mass. This procedure is implemented in an easy-to-use tool which can be applied to interpret present and future observations.

The paper is structured as follows: in Section~\ref{sec:methods} we discuss the initial conditions for the collisions, the interior models used to obtain the fragment radius and the suite of simulations. The results are presented and discussed Section~\ref{sec:results}. Finally, we give a summary and conclusions in Section~\ref{sec:summary}.

\begin{table}
    \centering
    \begin{tabular}{llll}
        \toprule
        Planet & Mass [$M_{\oplus}$] & Radius [$R_{\oplus}$] & $Z_{\textnormal{Fe}}$\\
        \midrule
        K2-38b & $7.3_{-1.0}^{+1.1}$ & $1.54\pm0.14$ & $0.63_{-0.36}^{+0.29}$\\
        Kepler-107c & $9.39\pm 1.77$ & $1.597\pm0.026$ & $0.68_{-0.16}^{+0.14}$\\
        K2-229b & $2.59\pm0.43$ & $1.164_{-0.048}^{+0.066}$& $0.67_{-0.32}^{+0.19}$\\
        L 168-9b & $4.60\pm 0.56$ & $1.390\pm0.090$ & $0.56_{-0.31}^{+0.25}$\\
        K2-106b & $8.36_{-0.94}^{+0.96}$ & $ 1.520\pm0.160$ & $0.74_{-0.36}^{+0.24}$ \\
        Kepler-80d & $3.70_{-0.60}^{+0.80}$ & $1.309_{-0.032}^{+0.036}$ & $0.63_{-0.36}^{+0.24}$\\
        K2-291b & $6.49\pm1.16$ & $1.589_{-0.072}^{+0.095}$ & $0.46_{-0.38}^{+0.23}$\\
        Kepler-406b & $6.35\pm1.40$ & $1.430\pm0.030$ & $0.73_{-0.25}^{+0.15}$ \\
        GJ 367b & $0.546\pm0.078$ & $0.718\pm 0.054$ & $0.82_{-0.32}^{+0.16}$\\
        \toprule
    \end{tabular}
    \caption{\textbf{Properties of the iron-rich exoplanets referenced in the paper.} The planet masses and radii are taken from the respective publication referenced in Section~\ref{sec:introduction}. The iron mass fractions $Z_{\textnormal{Fe}}$ are obtained from interior modelling as described in Section~\ref{sec:methods_interior_models}.}
    \label{tab_ExoplanetProperties}
\end{table}

\section{Methods}
\label{sec:methods}
We perform a large suite of 168 simulations considering different target and impactor masses and impact velocities. The mass of the target is \SI{10.5}{\Mearth}, \SI{16.55}{\Mearth} or \SI{19.04}{\Mearth} and the impactor mass is \SI{1}{\Mearth}, \SI{2}{\Mearth}, \SI{5}{\Mearth} or \SI{10}{\Mearth}. As the most extreme case we also investigate impacts with an impactor-to-target mass ratio of one, e.g., a \SI{10.5}{\Mearth} target colliding with a \SI{10.5}{\Mearth} impactor. The impacts are head-on, i.e., occur at an impact parameter $b=0$. Such events are expected to be the most erosive, they therefore provide a lower limit on the impact velocity required to obtain a certain composition. The impact velocities range from two to six times the mutual escape velocity which is given by: 
\begin{equation}
    v_{\textnormal{esc}}= \sqrt{\frac{2G(M_{\textnormal{tar}}+M_{\textnormal{imp}})}{R_{\textnormal{tar}}+R_{\textnormal{imp}}}}, 
\end{equation}
where $M_\textnormal{tar}$, $M_\textnormal{imp}$, $R_\textnormal{tar}$ and $R_\textnormal{imp}$ are the mass and radius of the target and the impactor, respectively, and $G$ is the gravitational constant.

\subsection{Impact simulations}
\label{sec:methods_impact_sim}
The impact simulations are performed using the Smoothed Particle Hydrodynamics (SPH) code \textsc{gasoline} \citep{Wadsley2004} with modifications for modelling giant impacts \citep{Reinhardt2017, Chau2018, Reinhardt2020, Meier2021}. The particle representations of the colliding planets are obtained using \textsc{ballic} \citep{Reinhardt2017, Chau2018}. We assume that the bodies are differentiated and have a chondritic composition consisting of a \SI{30}{\percent} by mass iron core surrounded by a \SI{70}{\percent} by mass rocky mantle. The thermal profile is adiabatic and the surface temperature of all models is set to \SI{1000}{\kelvin}. Due to noise inherent in SPH a particle’s density and internal energy are not consistent with the imprinted adiabatic thermal profile. We therefore first calculate all particles' densities with \textsc{gasoline} and then correct their internal energies so that all particles are on the same isentrope. The resulting models exhibit little noise because the particles are close to their equilibrium state but we still evolve each model in isolation in order to further relax the root mean square velocities to below \SI{1}{\meter\per\second}, which is a hundred fold reduction in the velocities. 
\subsubsection{Equations of State}
\label{sec:methods_impact_sim_eos}
To model the materials of the colliding planets we use the iron \citep{Stewart2020} and forsterite \citep{Stewart2019} equation of state (EOS) constructed from M-ANEOS (ANalytic Equation Of State) \citep{Thompson1974, Melosh2007, Thompson2019}. These EOS are an improvement over the standard M-ANEOS and ANEOS because the thermal part of the free energy is more accurately modelled in the liquid phase \citep{Stewart2020a} providing a significantly better fit to state-of-the-art high-pressure shock experiments \citep{Davies2020}. This improvement is crucial when modelling impacts involving bodies of more than \SI{10}{\Mearth} due to the huge peak pressure (and temperature) reached during shock compression.

\subsubsection{Resolution}
\label{sec:methods_impact_sim_res}
Recently, \citet{Meier2021} demonstrated that the amount of mass stripped in a collision shows little resolution dependence when ANEOS is used and has already converged at a resolution of $10^5$ particles. We therefore choose a resolution of $2 \times 10^5$ particles for the \SI{16.55}{\Mearth} body and adapt the resolution of all other bodies so that the particle mass is constant ensuring numerical stability. In all the simulations the target is therefore resolved with at least $10^5$ particles, which is sufficient to ensure that the bound mass has converged.

\subsubsection{Analysis}
\label{sec:methods_impact_sim_analysis}
The remaining bound mass, as well as the critical specific impact energy, \QRD, required to disrupt and gravitationally disperse half of the total colliding mass are determined as in \citet{Meier2021}. The mass of the gravitationally bound group of particles after the impact is calculated using the group finder \textsc{skid}\footnote{The source code is available at: \url{https://github.com/N-BodyShop/skid}.}. We find that the results are rather insensitive to the choice of parameters of the group-finding algorithm and that the fragment mass has converged in all simulations with \texttt{nSmooth} = 3200 and \texttt{tau} = 0.06. In a few cases \textsc{skid} finds multiple small fragments that did not merge with the largest fragment until the end of the simulation (\SI{53.1}{\hour} after impact). In these cases we first check whether any of the smaller fragments are bound to the largest fragment. We then calculate the bound mass and see if any other fragments are bound with respect to the aggregate's centre of mass. Those are again added to the bound mass and the procedure is repeated until no additional bound fragments are found. This algorithm usually converges after one or two steps since the largest fragment contains most of the total bound mass and therefore dominates the gravitational potential of the bound material. For a given target and impactor mass, the specific impact energy $Q_\textnormal{R}$ is determined for all impact velocities and is given by: 
\begin{equation}
    Q_\textnormal{R}=\frac{1}{2} \frac{\mu v_{\textnormal{imp}}^2}{M_{\textnormal{tot}}}, 
\end{equation}
where $M_{\textnormal{tot}}$ is the total colliding mass, $v_{\textnormal{imp}}$ the impact velocity and $\mu$ the reduced mass. The critical specific impact energy for catastrophic disruption \QRD is then calculated by linear interpolation between the two data points that bracket the specific impact energy where exactly half of the total colliding mass remains bound.

\subsection{Interior models}
\label{sec:methods_interior_models}
After the collision we determine the bound mass as described above and assume that the planet cools down to the pre-impact surface temperature of \SI{1000}{\kelvin}. This temperature is comparable to the surface temperatures of the close-in iron-rich exoplanets in our sample. The surface temperature of the planet depends on the stellar type and age, the distance of the planet from the host star, and the planetary thermal evolution. We find that for values between \SI{300}{\kelvin} and \SI{1500}{\kelvin} the impact of the temperature on the inferred radius is very small ($< \SI{1}{\percent}$). For each fragment's mass and composition we determine the radius using the interior model of \citet{Otegi2020}, which assumes a pure iron core and a silicate mantle. The EOS used to model the iron core is taken from \cite{Hakim2018}. The constituents of the mantle are assumed to be SiO$_2$, MgO, CaO, Al$_2$O$_3$ and NaO$_2$. The stable minerals at a given pressure, temperature and composition are calculated by Gibbs energy minimisation using the \textsc{Perple\_X} code presented in \cite{Connolly2009} and the thermodynamical model of \cite{Stixrude2011}. We assume an adiabatic thermal profile for both the core and the mantle and that the abundances of the mantle constituents are identical to the Earth's mantle which is consistent with the EOS we use for the impact simulations. For more details about the model we refer the reader to \cite{Otegi2020} and references therein. 

In order to verify consistency between these models and the ones uses for the impact simulations we also determine each post-impact fragment's radius with \textsc{ballic} and M-ANEOS using the same material parameters as for the initial conditions of the impact simulations. We find that for most masses and compositions, the radii of the resulting bodies agree within \SI{2}{\percent} for the two methods. Only for very small masses and large iron mass fractions the agreement is worse but is still within \SI{5}{\percent}. Generally, the models obtained using \textsc{ballic} and M-ANEOS slightly overestimate the planet's radius compared to the interior model described above. Considering that M-ANEOS is designed for modelling impact simulations and therefore aims at covering a large range of densities and temperatures using analytic expressions for the free energy, we consider this an excellent agreement. In any case, the dependence of the impact simulation outcome on the precise interior model (considering mineral level composition) is negligible.

\section{Results and Discussion}
\label{sec:results}
\subsection{Simulation result}
\label{sec:results_sim}
\begin{figure}
    \includegraphics[width=\columnwidth]{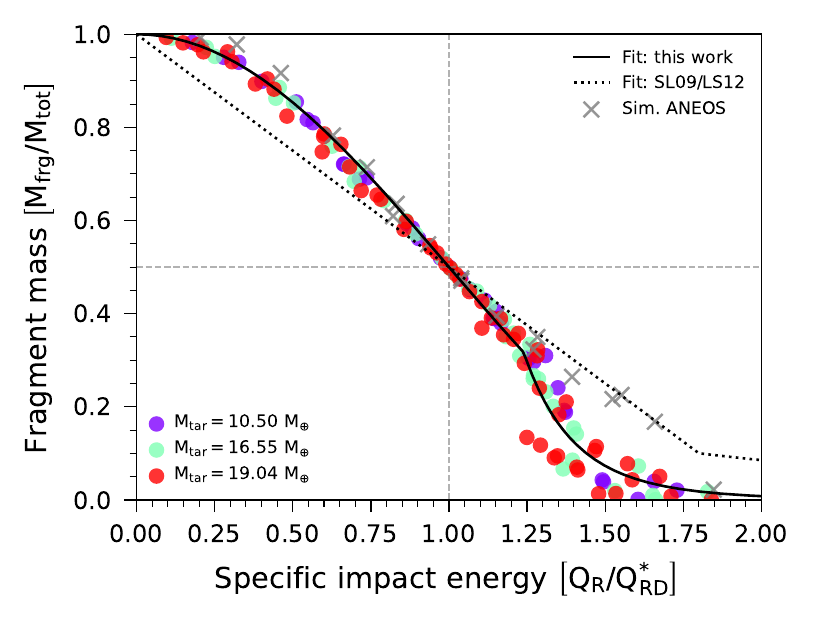}
    \caption{\textbf{The mass of the largest fragment for different specific impact energies.} The circles represent the simulation data and each colour corresponds to the target's pre-impact mass (violet: \SI{10}{\Mearth}, cyan: \SI{16.55}{\Mearth} and red: \SI{19.04}{\Mearth}). The critical specific impact energy \QRD is determined from the simulation data for each combination of target and impactor mass. Despite the large difference in the total colliding mass ranging from \SI{10}{\Mearth} to \SI{40}{\Mearth} the normalised fragment mass and specific impact energies show little scatter and are in excellent agreement. We observe that for $Q_\textnormal{imp}/Q_\textnormal{RD}^{*} \approx 1.26$ the bound mass rapidly decreases and follows a power law. While the scaling law presented in this work (solid black line) provides an excellent fit to the simulation data the results deviate significantly both below and above \QRD from the behaviour predicted by the scaling laws of SL09 and LS12. If ANEOS is used for identical initial conditions (grey crosses), substantially less material is stripped due to the lack of an accurate treatment of the liquid phase (for details see Section~\ref{sec:results_sim}).}
    \label{fig:mass_frg_vs_q_imp}
\end{figure}
\begin{figure}
    \includegraphics[width=\columnwidth]{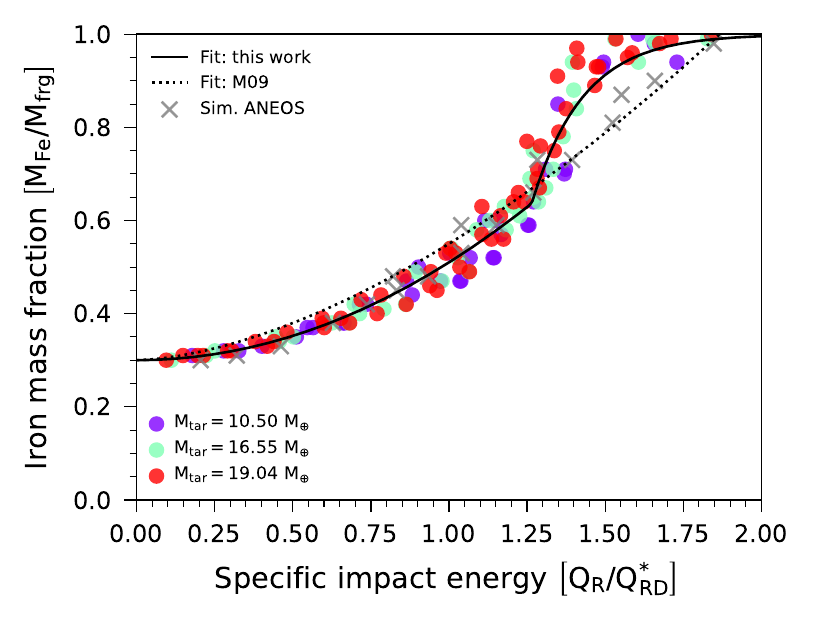}
    \caption{\textbf{The iron mass fraction of the largest fragment for different specific impact energies.} The circles represent the simulation data and the colours correspond to the target's pre-impact mass (violet: \SI{10}{\Mearth}, cyan: \SI{16.55}{\Mearth} and red: \SI{19.04}{\Mearth}). As in Figure~\ref{fig:mass_frg_vs_q_imp} the critical specific impact energy \QRD is determined from the simulation data. The solid black line is our best fit and the dotted black line the scaling law from M09. At lower impact energies the scaling law from M09 overestimates the amount of mantle material lost due to mantle stripping. The largest difference between the two scaling laws is at very high impact energies due to the strong disruption above $Q_\textnormal{imp}/Q_\textnormal{RD}^{*} \approx 1.26$ observed in our simulation. Furthermore, our scaling laws correctly extrapolate to $Z_\textnormal{Fe}=1$ in the limit of super-catastrophic disruption. These differences have profound implications for the impact velocity required to obtain a certain fragment mass and composition (see Section~\ref{sec:results_derived_scaling_laws}).}
    \label{fig:iron_mass_frac_vs_q_imp}
\end{figure}

Figure~\ref{fig:mass_frg_vs_q_imp} shows the mass of the largest fragment versus the specific impact energy for all our simulations. Generally the impact conditions required to strip a certain amount of the mantle depend on the target mass and the impactor-to-target mass fraction $\gamma = M_{\textnormal{imp}}/M_{\textnormal{tar}}$. However, once the bound mass is normalised to the total colliding mass and the specific impact energy to \QRD (determined from the simulation data as described in Section~\ref{sec:methods_impact_sim_analysis}) the resulting dimensionless quantities are in excellent agreement for all simulations. Only in the case of a super-catastrophic disruption of the colliding bodies the normalised bound mass is more noisy as the fragment is resolved with only a few thousand particles. At low normalised specific impact energies very little mass is stripped in the collision. Around the specific impact energy for critical disruption \QRD we observe a linear dependence of the bound mass on the impact energy as reported in SL09 and M09. However, in the critical disruption regime, around $Q_{\textnormal{imp}}/Q_{\textnormal{RD}^{*}} \gtrsim 1.26$, we find that the bound mass rapidly decreases and deviates from the linear behaviour reported in \citet{Stewart2009} (hereafter SL09) and M09 following a power law. In \citet{Leinhardt2012} (hereafter LS12) such a behaviour was observed for super-catastrophic disruption, which was defined as $M_{\textnormal{frg}}/M_{\textnormal{tot}} \lesssim 0.1$. Previous work was limited to substantially smaller target--impactor masses as well as lower impact velocities. Therefore, a reason for the difference in behaviour could be the extreme impact energies probed in our simulations; energies for which the peak shock pressures and temperatures are very high. Another difference to previous work is that older versions of ANEOS and M-ANEOS were used. These EOS model the thermal part of the free energy in the liquid phase as a solid and therefore disagree with data from high-pressure shock experiments \citep{Davies2020}. The modified version of M-ANEOS of \citet{Stewart2019} used in this work substantially improves the treatment of the liquid phase and provides a much better match to experiments in the region relevant for such energetic impacts. In order to test this hypothesis we perform impact simulations with ANEOS (the material parameters for iron and dunite can be found in \citealt{Meier2021}) of the two most extreme cases: a \SI{10.5}{\Mearth} target colliding with a \SI{1}{\Mearth} impactor and a \SI{19.04}{\Mearth} target colliding with a \SI{19.04}{\Mearth} impactor for the entire range of impact velocities investigated in this paper. We determine \QRD for both cases and find that generally less mass is stripped compared to the M-ANEOS simulations. This difference is most pronounced for large specific impact energies where the normalised bound mass shows a linear dependency on impact energy (see the grey crosses in Figure~\ref{fig:mass_frg_vs_q_imp}). Given that the impact conditions are identical to the ones of the M-ANEOS simulations we conclude that indeed the more accurate treatment of the liquid phase in M-ANEOS is responsible for the different behaviour at very high impact energies. This again demonstrates that the EOS plays a key role in faithfully modelling GI \citep{Stewart2019,Meier2021}. 

In order to test a possible connection between the transition to rapid mass loss and the initial core mass fraction we perform an additional set of simulations where we assume an iron mass fraction of $Z_{\textnormal{Fe}} = 0.5$. The initial conditions are identical to the other simulations, and we again consider impacts between a \SI{10}{\Mearth} target and a \SI{1}{\Mearth} impactor as well as collisions between two bodies of \SI{19.04}{\Mearth}. We find that the transition to rapid mass loss indeed depends  on the initial core mass fraction. The value of $Q_{\textnormal{imp}}/Q_{\textnormal{RD}^{*}}$ at which the transition occurs depends on the EOS of iron as well as the initial core mass fraction. However, this dependency is limited to energies above the threshold for rapid mass loss while the behaviour at lower specific impact energies does not depend on the initial core mass fraction. We also find that the bound mass exhibits substantially more noise at impact energies above the threshold for rapid mass loss because the fragments are resolved with fewer particles. Therefore the behaviour in this regime could also be resolution-dependent. Investigating these two aspects is beyond the scope of this paper and we hope to address it in future work.

We present the corresponding iron mass fractions in Figure~\ref{fig:iron_mass_frac_vs_q_imp}. At low impact energies, i.e., $Q_{\textnormal{imp}}/Q_{\textnormal{RD}^{*}} \leq 0.5$, little mass is stripped and therefore the deviation of the iron mass fraction from the pre-impact value is very small. For impact energies higher than $Q_{\textnormal{imp}}/Q_{\textnormal{RD}^{*}} \approx 1.26$ on the other hand the iron mass fraction massively increases as the bound mass decreases following a power law. These findings have profound implications for the stripping of a planet's mantle due to a giant impact and therefore for the likelihood of the formation and observation of iron-rich exoplanets. First, since relatively large specific impact energies are required one would expect that most GI are insufficient to produce super-Mercuries. Second, in order to obtain a iron-rich massive planet, aside from large impact velocities, the initial (pre-impact) mass must be extremely high since much of the mass is lost in the critical disruption regime. This is consistent with the fact that iron-rich exoplanets are rather rare and few with masses $> \SI{8}{\Mearth}$ have been observed.

Finally, except for the most disruptive collisions, we find that a fraction of the impactor remains bound. The mass fraction of impactor material in the post-impact planet $f_{\textnormal{imp}}$ increases with increasing impactor mass and can be up to \SI{50}{\percent} in the case of equal mass collisions, as expected by the symmetry of the impact. For small impactor-to-target mass ratios, i.e., $\gamma = M_{\textnormal{imp}}/M_{\textnormal{tar}} < 0.2$ the mass fraction of impactor material decreases with increasing impact velocity for a given impactor mass. For the larger impactor masses we find that $f_{\textnormal{imp}}$ is approximately equal to the impactor mass divided by the total colliding mass and is independent of the impact velocity. A mantle stripping impact can therefore leave an isotopic fingerprint if the impactor accreted from a very different region of the proto-planetary disk and the isotopic difference in composition is preserved after the collision. Furthermore, the amount of impactor material in the post-impact planet {\em in principle} allows one to predict the impactor mass and therefore could provide constraints on the impact conditions.

\subsection{Scaling laws}
\label{sec:results_sl}
\subsubsection{Fragment mass}
\label{sec:results_sl_mass_frg}
As in previous studies we derive scaling laws from the simulation data. For the mass of the largest fragment SL09 found that an expression of the form
\begin{equation}
    \frac{M_\textnormal{frg}}{M_\textnormal{tot}} = -\frac{1}{2} \left(\frac{Q_{\textnormal{R}}}{Q_{\textnormal{RD}}^{*}} - 1 \right) + \frac{1}{2}
\end{equation}
best fits their simulation data. While this fit is appropriate for the relatively low mass and velocity collisions investigated in their paper it does not accurately represent our data as shown in Figure~\ref{fig:mass_frg_vs_q_imp}. At specific impact energies below \QRD the bound mass is underestimated because the massive targets considered in our study are more resilient to erosion. Likewise, at specific impact energies above \QRD the bound mass is strongly overestimated because the bound mass decreases rapidly following a power law beyond $Q_{\textnormal{R}}/Q_{\textnormal{RD}}^{*} \approx 1.26$ as proposed in LS12 for super-catastrophic disruption.

We therefore propose a new scaling law for the mass of the largest fragment,
\begin{equation}
    \frac{M_\textnormal{frg}}{M_\textnormal{tot}} = \begin{dcases}
    \frac{1}{2}\cos{ \left(\frac{\pi}{2} \frac{Q_{\textnormal{R}}}{Q_{\textnormal{RD}}^{*}}\right)} + \frac{1}{2} & \text{if } \frac{Q_{\textnormal{R}}}{Q_{\textnormal{RD}}^{*}} \le q_{\textnormal{sc}},\\
        b_{\textnormal{M}} \left( \frac{Q_{\textnormal{R}}}{Q_{\textnormal{RD}}^{*}} \right)^{\eta} & \text{otherwise},
    \end{dcases}
    \label{eqn:mass_frg_vs_q_imp}
\end{equation}
where,
\begin{equation}
    b_{\textnormal{M}} = \frac{ \cos{\left( \frac{\pi}{2} q_{\textnormal{sc}} \right)} + 1}{2 q_{\textnormal{sc}}^{\eta}},
\end{equation}
is given by the requirement of continuity between the two functions. The coefficients that best describe our results are $q_{\textnormal{sc}}=1.2645$ and $\eta = -8.1214$. The coefficient $q_{\textnormal{sc}}$ corresponds to the specific impact energy at which the bound mass starts to decrease following a power law and therefore marks the transition to extreme mass loss in collisions which we observe earlier than in prior simulations that involved lower mass bodies. Furthermore, the value of $\eta$ we find is smaller than the $\eta \approx -1.5$ that was reported for super-catastrophic disruption in LS12.

\subsubsection{Iron mass fraction}
\label{sec:results_sl_iron_mass_fraction}
For normalised specific impact energies $Q_{\textnormal{R}}/Q_{\textnormal{RD}}^{*}$ below $q_\textnormal{sc}$ the iron mass fraction of the remaining bound mass follows a power law as proposed in M09. At larger normalised specific impact energies a power law provides a poor fit to our simulation data and severely underestimates the iron mass fraction due to the rapid mass loss beyond $Q_{\textnormal{R}}/Q_{\textnormal{RD}}^{*} \approx 1.26$ (see Figure~\ref{fig:iron_mass_frac_vs_q_imp}). Additionally, the predicted iron mass fraction can exceed $Z_\textnormal{Fe}=1$ for very disruptive collisions which is clearly unphysical.

We find that a function of the form,
\begin{equation}
    Z_\textnormal{Fe} = \begin{dcases}
        Z_\textnormal{Fe,i} + a_\textnormal{Fe} \left(\frac{Q_{\textnormal{R}}}{Q_{\textnormal{RD}}^{*}}\right)^{b_\textnormal{Fe}} & \text{if } \frac{Q_{\textnormal{R}}}{Q_{\textnormal{RD}}^{*}}\le q_{\textnormal{sc}},\\
        1 - a_\textnormal{Fe,sc} \exp{\left[-{b_\textnormal{Fe,sc} \left(\frac{Q_{\textnormal{R}}}{Q_{\textnormal{RD}}^{*}}- q_{\textnormal{sc}}\right)}\right]} & \textnormal{otherwise,}
    \end{dcases}
    \label{eqn:iron_mass_frac_vs_q_imp}
\end{equation}
best fits our data. For our simulations the initial (pre-impact) iron mass fraction is $Z_\textnormal{Fe,i}=0.3$. For lower impact energies the fitting coefficients $a_\textnormal{Fe} = 0.2099$ and $b_\textnormal{Fe} = 2.1516$ differ from the ones found in M09 ($a_\textnormal{Fe}=0.25$ and $b_\textnormal{Fe}=1.65$) corresponding to the lower iron mass fraction and change in composition found in our simulations. For impact energies above $q_{\textnormal{sc}}$ the coefficient
\begin{equation}
    a_\textnormal{Fe,sc} = 1 - \left( Z_\textnormal{Fe,i} + a_\textnormal{Fe} q_{sc}^{b_\textnormal{Fe}}\right)
\end{equation}
is determined by requiring continuity of $Z_\textnormal{Fe}$ at $q_{\textnormal{sc}}$ and $b_\textnormal{Fe,sc} = 6.4890$. The iron mass fraction predicted by this scaling law not only is in excellent agreement with our simulation data but also extrapolates to $Z_\textnormal{Fe}=1$ for very large specific impact energies. The scaling laws derived from this equation in Section~\ref{sec:results_derived_scaling_laws} never provide unphysical values (i.e., $Z_\textnormal{Fe} > 1$).

\subsubsection{Critical specific impact energy for catastrophic disruption}
\label{sec:results_sl_qrd}
\begin{figure}
    \includegraphics[width=\columnwidth]{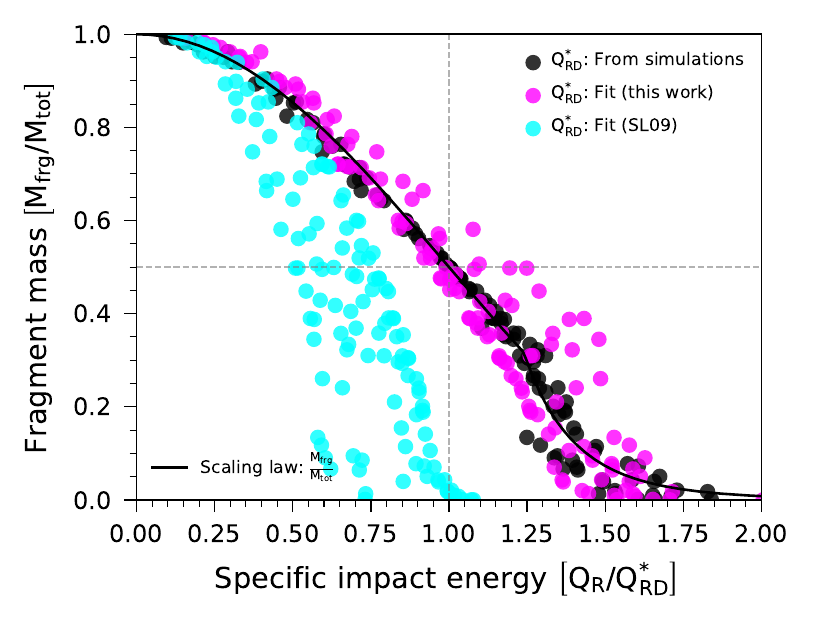}
    \includegraphics[width=\columnwidth]{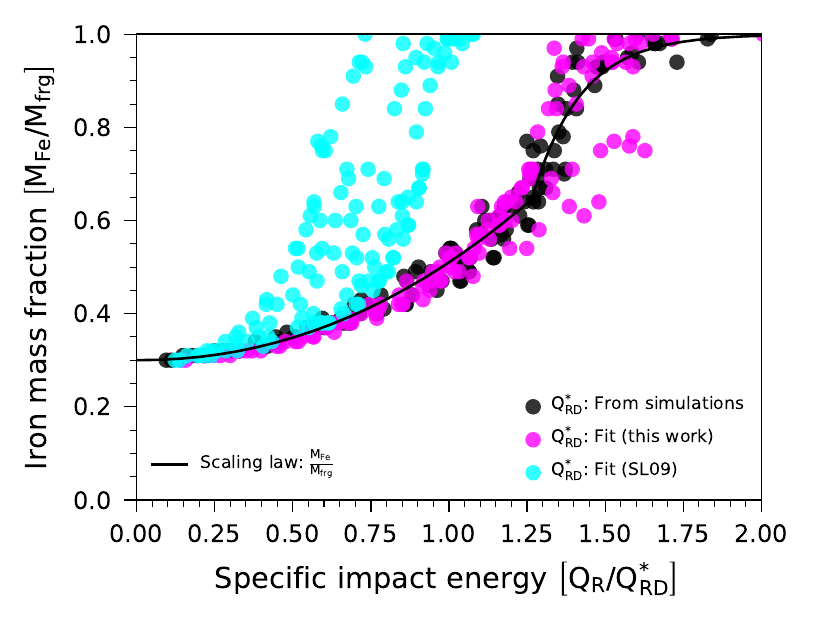}
    \caption{\textbf{The impact of different fitting coefficients for calculating the critical specific impact energy from scaling laws on the normalised mass (top panel) and iron mass fraction (bottom panel) of the largest fragment.} The circles represent the simulation data and the colours the different values of \QRD used to normalise the specific impact energy (black: from the simulations, magenta: scaling law with fitting parameters derived in this work and cyan: scaling law from SL09). If \QRD is calculated from the scaling law with the fitting parameters presented in SL09 the values deviate by up to a factor of $\sim 2.5$. This introduces large scatter in the data when normalising the specific impact energy to the value derived from the scaling laws. Furthermore, catastrophic disruption occurs already at $\sim0.5$ \QRD and for specific impact energies larger than \QRD no mass remains bound. If the updated fitting parameters derived in the present work are used, \QRD agrees within $\leq\SI{33}{\percent}$ with the value derived from the simulations and the scatter in the data is substantially reduced. In turn catastrophic and super-catastrophic disruption are in much better agreement with simulation data.}
    \label{fig:effect_of_qrd_scalinglaw}
\end{figure}

Finally, we compare the values of \QRD calculated from the velocity-dependent catastrophic disruption criterion proposed in SL09 (based on dimensional analysis following \citet{Housen1990, Housen1999}) to the values inferred from our simulations. SL09 find that in the gravity dominated regime
\begin{equation}
    Q_{\textnormal{RD}}^{*} = q_g R_{C1}^{3 \mu} v_{\textnormal{imp}}^{2-3\mu}
    \label{eqn:scaling_law_qrd}
\end{equation}
where $R_{C1}$ is the radius of a sphere containing the total colliding mass with a mean density of \SI{1}{\gram\per\centi\meter\tothe3} with $q_g = 10^{-4}$ (in cgs units) and $\mu=0.4$ is in good agreement with their simulation data. However, these results cannot be extrapolated to collisions involving orders of magnitude more massive bodies and the calculated values of \QRD differ by a factor of $\sim 2.5$ from the values derived from our simulations. Given the huge range of extrapolation this disagreement is not surprising. When normalising \QR from our simulations to \QRD calculated with these fitting coefficients one introduces large scatter and shifts the transition to super-catastrophic disruption to below $Q_{\textnormal{R}}/Q_{\textnormal{RD}}^{*} = 1$ (see Figure~\ref{fig:effect_of_qrd_scalinglaw}). This is clearly unrealistic and results in large errors when deriving additional equations based on these scaling laws. We therefore determine the fitting coefficients from our simulation data. For $q_g=\SI{1.4679e-06}{}$ (cgs) and $\mu=\SI{0.6164}{}$ the values of \QRD calculated from Equation~\ref{eqn:scaling_law_qrd} are in very good agreement with our results and the largest relative error is smaller than \SI{33}{\percent}. This substantially reduces the scatter in Figure~\ref{fig:effect_of_qrd_scalinglaw} and allows a reliable prediction of \QRD without requiring a large suite of simulations.

\begin{figure}
    \includegraphics[width=\columnwidth]{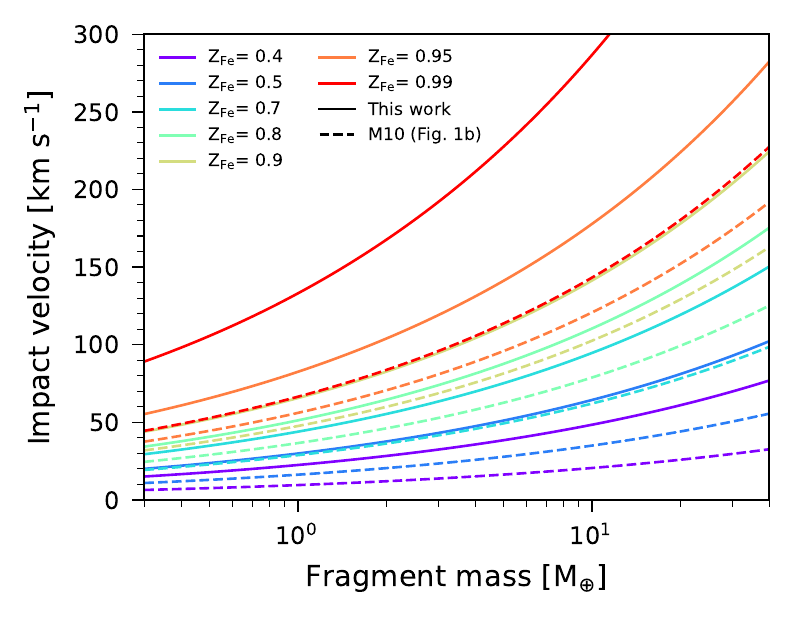}
    \caption{\textbf{The impact velocity versus the mass of the largest fragment for different iron mass fractions predicted by scaling laws.} Shown are our results (solid lines) and the ones from Figure 1b in M10 (dashed lines). The target-to-impactor mass ratio is fixed to $\gamma=1$ here, and the colours represent different iron mass fractions (see legend). For a given fragment mass the impact velocity required to obtain a desired iron mass fraction is substantially higher than predicted by M10. In order to obtain a \SI{10}{\Mearth} pure iron body an impact velocity of about \SI{300}{\kilo\meter\per\second} corresponding to three times the Keplerian velocity of a planet orbiting a Sun like star at \SI{0.1}{\astronomicalunit} is required thus making the collisional formation of such bodies very challenging.}
    \label{fig:sl_vel_imp_vs_frg_mass}
\end{figure}
\begin{figure}
    \includegraphics[width=\columnwidth]{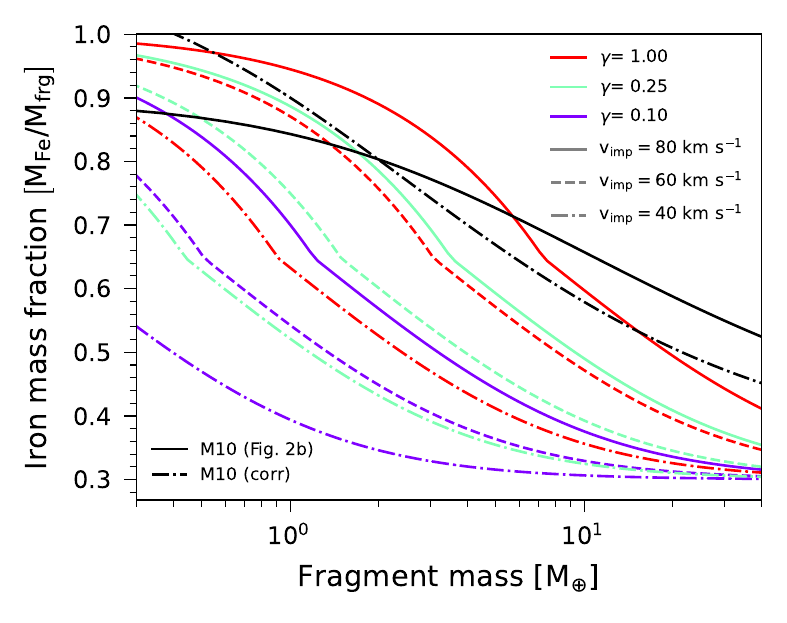}
    \caption{\textbf{The iron mass fraction versus the mass of the largest fragment for different impact velocities and target-to-impactor mass ratios predicted by the scaling laws derived in this work.} The colours represent different impact velocities and the line styles different target-to-impactor mass ratios (see legend). The earlier transition from normal to super-critical disruption behaviour observed in our simulations is affecting the slope of the curves and results in a kink where the function is continuous but not differentiable. We also plotted the results from M10 for an impact velocity of \SI{80}{\kilo\meter\per\second} and $\gamma=1$. The curve shown in Figure 2b of M10 (solid black line) under and over predicts the iron mass fraction by $\sim\SI{10}{\percent}$ at small and larger fragment masses respectively. Even if the curve is corrected for typos and other errors (dot-dashed black line, see Appendix~\ref{appendix:discussion_results_m10} for details) the predicted iron mass fraction is very different from our results. This affects the minimum radius of a planet due to collisional stripping of the mantle derived in M10 (see Figure~\ref{fig:sl_radius_vs_frg_mass}).}
    \label{fig:sl_iron_mass_frac_vs_frg_mass}
\end{figure}
\begin{figure}
    \includegraphics[width=\columnwidth]{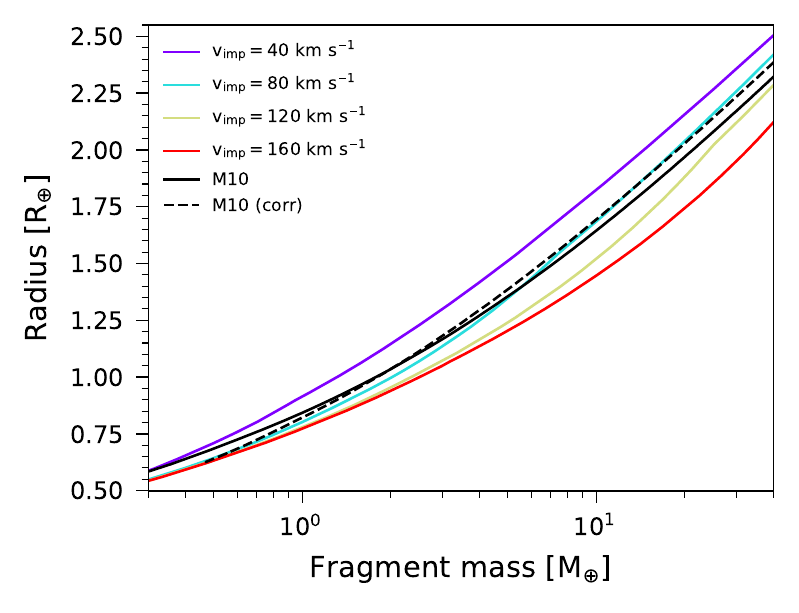}
    \includegraphics[width=\columnwidth]{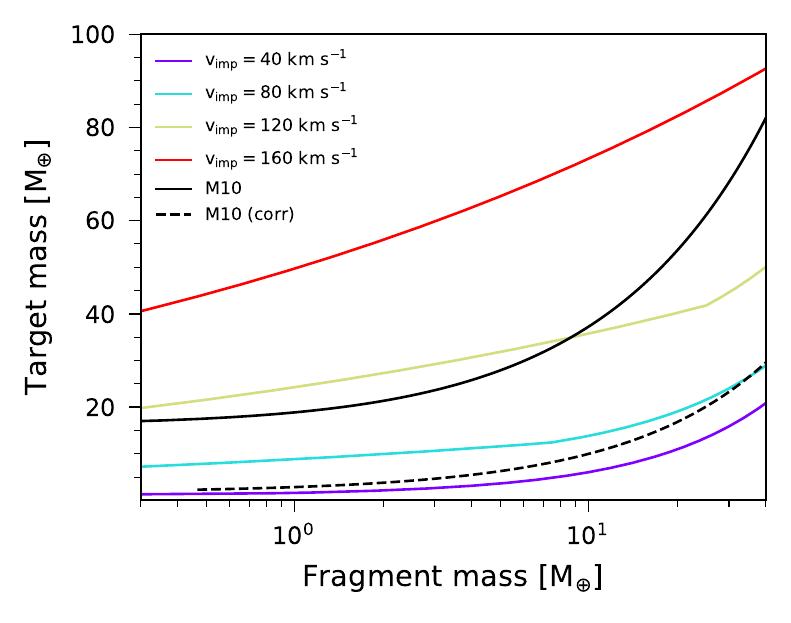}
    \caption{\textbf{The inferred radius (top) and target mass (bottom) versus the mass of the largest fragment for different impact velocities predicted by our scaling laws.} The colours represent different impact velocities and the black curves (solid: iron mass fraction in Figure 2b of M10, dashed: corrected as described in Appendix~\ref{appendix:discussion_results_m10}) correspond to the minimum radius predicted in M10 assuming an impact velocity of \SI{80}{\kilo\meter\per\second}. The impactor-to-target mass ratio is one for all curves. The largest difference in radius between M10 and this work is \SI{6.5}{\percent} which is close to the maximum uncertainty in radius in \citet{Otegi2020}. Future missions are expected to substantially reduce the uncertainty in the planetary radius making such differences relevant when interpreting observations. For a given fragment mass higher impact velocities require higher initial target masses, e.g., for an impact velocity of \SI{160}{\kilo\meter\per\second} the target has to have a mass between \SI{40}{\Mearth} and \SI{93}{\Mearth}. Once the equations of M10 are corrected for all errors, our scaling laws require larger initial target masses than predicted in M10 due to the stronger disruption of the bodies found in our simulations.}
    \label{fig:sl_radius_vs_frg_mass}
\end{figure}
\begin{figure*}
    \includegraphics[width=0.75 \linewidth]{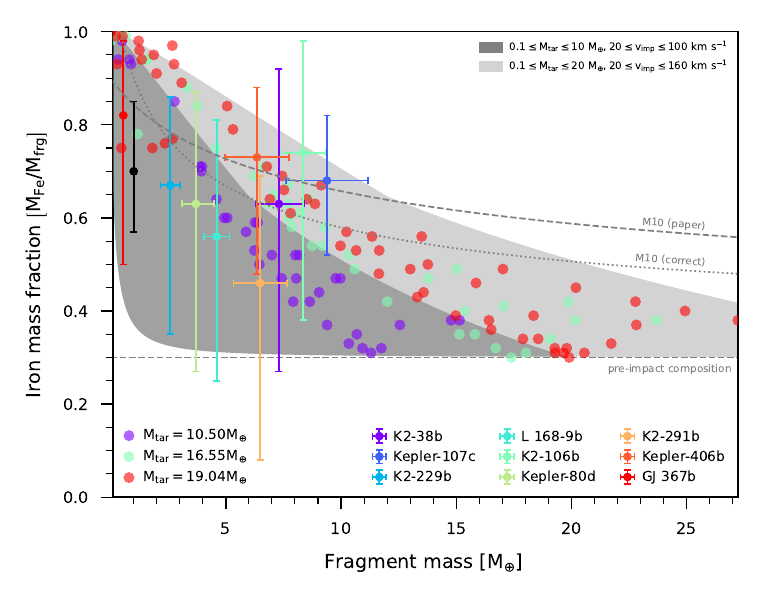}
    \caption{\textbf{The post-impact iron mass fraction for different fragment masses.} The circles represent the simulation data and each colour corresponds to the target's pre-impact mass (violet: \SI{10}{\Mearth}, cyan: \SI{16.55}{\Mearth} and red: \SI{19.04}{\Mearth}). The most iron-rich and massive fragments are obtained from collision onto a \SI{19.04}{\Mearth} target so in order to obtain a metal-rich, massive planet large initial (pre-impact) masses are required. The dashed and dotted grey curves correspond to the original and corrected maximum iron content due to collisional stripping of a planet's mantle derived in M10 (see Appendix~\ref{appendix:discussion_results_m10} for details). The shaded grey area mark all possible outcomes for a restricted impact parameter space (dark grey:  $\SI{0.1}{\Mearth} \leq M_\textnormal{tar} \leq \SI{10}{\Mearth}$ and $\SI{20}{\kilo\meter\per\second} \leq v_\textnormal{imp} \leq \SI{100}{\kilo\meter\per\second}$, light grey:  $\SI{0.1}{\Mearth} \leq M_\textnormal{tar} \leq \SI{20}{\Mearth}$ and $\SI{20}{\kilo\meter\per\second} \leq v_\textnormal{imp} \leq \SI{160}{\kilo\meter\per\second}$ both with $\gamma = 1$).  The iron-rich exoplanets's (see legend for details) compositions are derived from interior modelling as described in Section~\ref{sec:methods_interior_models} and are in good agreement with masses and compositions obtained from simulations. Their large iron mass fractions therefore could be the result of a GI that stripped a fraction of the mantle. The black marker shows the composition of a hypothetical super-Mercury with a mass of $\SI{1}{\Mearth}$ and $Z_{\textnormal{Fe}}=0.7$ and uncertainties of \SI{10}{\percent} and \SI{3}{\percent} on the mass and radius,  respectively, to forecast the expected accuracy of future observations}. 
    \label{fig:composition_vs_mass_frg}
\end{figure*}
\begin{figure*}
    \includegraphics[width=0.75 \linewidth]{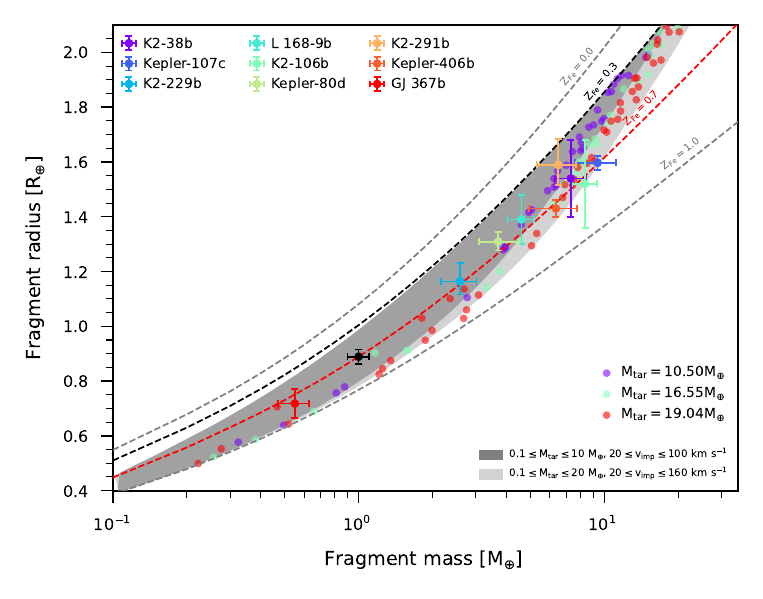}
    \caption{\textbf{The post-impact radius for different fragment masses corresponding to the iron mass fractions shown in Figure~\ref{fig:composition_vs_mass_frg}.} The circles represent the simulation data and each colour corresponds to the target's pre-impact mass (violet: \SI{10}{\Mearth}, cyan: \SI{16.55}{\Mearth} and red: \SI{19.04}{\Mearth}). The dashed grey lines present a pure-rock (top curve) and pure-iron (bottom curve) body. The pre-impact composition is shown in black (dashed line) and the dashed red line corresponds to $Z_{\textnormal{Fe}}=0.7$, i.e., a Mercury-like composition. The shaded grey regions mark different impact parameter spaces (dark grey: $\SI{0.1}{\Mearth} \leq M_\textnormal{tar} \leq \SI{10}{\Mearth}$ and $\SI{20}{\kilo\meter\per\second} \leq v_\textnormal{imp} \leq \SI{100}{\kilo\meter\per\second}$, light grey: $\SI{0.1}{\Mearth} \leq M_\textnormal{tar} \leq \SI{20}{\Mearth}$ and $\SI{20}{\kilo\meter\per\second} \leq v_\textnormal{imp} \leq \SI{160}{\kilo\meter\per\second}$ both with $\gamma=1$). Shown are also the masses and radii of metal-rich exoplanets which are all in good agreement with our simulation data. The black marker shows a hypothetical super-Mercury with a mass of $\SI{1}{\Mearth}$ and $Z_{\textnormal{Fe}}=0.7$ and uncertainties of \SI{10}{\percent} and \SI{3}{\percent} on the mass and radius respectively to forecast future observations.}
    \label{fig:radius_vs_frg_mass}
\end{figure*}

\subsection{Derived scaling laws}
\label{sec:results_derived_scaling_laws}
The above scaling laws (Equation~\eqref{eqn:mass_frg_vs_q_imp},~\eqref{eqn:iron_mass_frac_vs_q_imp} and~\eqref{eqn:scaling_law_qrd}) can be combined in a similar way as in M10 to obtain additional scaling laws. Due to the more complex nature of our fitting functions it is more convenient to use a numerical root finder (SciPy: \citet{Virtanen2020}) in place of an analytic solution as in M10. The resulting scaling laws allow one to predict the conditions, i.e., impact velocity, impactor-to-target mass ratio and total colliding mass, required to obtain a certain post-impact fragment mass and composition without performing a large suite of simulations.

Figure~\ref{fig:sl_vel_imp_vs_frg_mass} shows the impact velocity required to obtain a desired fragment mass and composition for our scaling laws as well as the ones presented in M10. The impactor-to-target mass ratio is fixed to $\gamma=1$ as such collisions strip the most material for a given impact velocity and total colliding mass and therefore provide a lower limit on the required relative velocities to obtain a given fragment mass and composition. As expected the required impact velocity increases with fragment mass and iron mass fraction. For example, obtaining a fragment of \SI{1}{\Mearth} with a Mercury-like composition, i.e., $Z_{\textnormal{Fe}}=0.7$, requires an impact velocity of \SI{44.14}{\kilo\meter\per\second}. This is comparable to the orbital velocity at Mercury's location. Generally, the impact velocities predicted by our scaling laws are substantially larger than the ones in M10. We also find that for large iron mass fractions the slopes of our curves are steeper, which means that for higher fragment masses, even larger impact velocities are required. The former is to some extent a result of the different fitting functions as well as the errors made in M10 when deriving the equations (see Appendix~\ref{appendix:discussion_results_m10} for details). Furthermore, as shown in Section~\ref{sec:results_sl_qrd} the fitting coefficients from SL09 used to estimate \QRD in M10 are derived for low mass collisions and are in poor agreement with values derived from high mass collision simulations. The normalised specific impact energy required for super-catastrophic disruption is therefore severely underestimated which in turn means that much lower impact velocities are needed to obtain a desired fragment mass and composition. The latter is a result of the simulations presented in M09 over-estimating the bound mass compared to our results and therefore under-estimating the iron mass fraction for a given specific impact energy. By pure coincidence, the curves for $Z_{\textnormal{Fe}}=0.90$ from our simulations and $Z_{\textnormal{Fe}}=0.99$ in M10 (their Figure 1b) are almost overlapping (see Figure~\ref{fig:sl_vel_imp_vs_frg_mass}).

Figure~\ref{fig:sl_iron_mass_frac_vs_frg_mass} shows the iron mass fraction for a given fragment mass for different impact velocities and impactor-to-target mass ratios. To compare with M10 we also added the curve for $\gamma=1$ and an impact velocity of \SI{80}{\kilo\meter\per\second} from their paper (both as shown in the paper and the correct version, see Appendix~\ref{appendix:discussion_results_m10} for details) which is used to determine the maximum iron mass fraction of an exoplanet due to collisional stripping of the mantle. For a given impact velocity and fragment mass larger impactor-to-target mass ratios result in more iron-rich bodies due to the higher specific impact energy. Likewise increasing the impact velocity for a given mass ratio $\gamma$ and fragment mass increases the iron mass fraction. The normalised specific impact energy $q_{\textnormal{sc}}$ where the fragment mass begins to abruptly decrease following a power law (see Equation~\ref{eqn:mass_frg_vs_q_imp}) results in a kink where the function is continuous but not differentiable. Furthermore, the results of M10 severely over-predict the iron mass fraction at large fragment masses. This is most pronounced for the curve as shown in Figure 2b of their paper. At the same time the iron mass fraction at small fragment masses is substantially smaller than in our simulations. If one corrects for the mistakes made when deriving the equation in M10 the curve is in better agreement with our results at large fragment masses but predicts an unphysically large iron content, i.e., $Z_{\textnormal{Fe}} > 1$, at small masses. Due to the correct extrapolation to $Z_{\textnormal{Fe}} = 1$ in Equation~\ref{eqn:iron_mass_frac_vs_q_imp} this is not the case for the scaling laws derived in this paper.

The fragment mass and composition obtained from these scaling laws can be used to estimate the radius after a collision from interior modelling as described in Section~\ref{sec:methods_interior_models}. The top panel of Figure~\ref{fig:sl_radius_vs_frg_mass} is an M-R diagram showing the radius versus fragment mass for different impact velocities assuming the fragment had time to cool to a surface temperature of \SI{1000}{\kelvin} after the collision. We consider fragment masses between \SI{0.3}{\Mearth} and \SI{40}{\Mearth} and the impact velocities vary between \SI{40}{\kilo\meter\per\second} and \SI{160}{\kilo\meter\per\second}. Again, the impactor-to-target mass ratio is $1$; correspondingly the curves shown in the figure have to be considered as a lower limit on the radius (see Section~\ref{sec:results_sl_iron_mass_fraction} for details). For a given fragment mass, larger impact velocities result in smaller radii since the remaining fragment contains more iron and is therefore more compact. The difference in radius between an impact velocity of \SI{40}{\kilo\meter\per\second} and \SI{160}{\kilo\meter\per\second} ranges between $8$ and $21$ percent and is most pronounced at a mass of \SI{8}{\Mearth}.

The cyan curve shows the radius of a fragment obtained at an impact velocity of \SI{80}{\kilo\meter\per\second}. This corresponds to the upper limit M10 assumed on the possible relative velocity when deriving the minimum radius of a fragment (and therefore an iron-rich exoplanet) due to collisional stripping. We also calculated the fragment radius for the masses and iron mass fractions obtained from the scaling laws presented in M10 (solid black line: including all mistakes, dashed black line: corrected equation) at identical impact conditions. This allows a direct comparison between the two studies and how differences could affect implications made from observations. As expected from Figure~\ref{fig:sl_iron_mass_frac_vs_frg_mass} the mass-radius curve inferred from the compositions shown in Figure 2b of M10 overestimates the fragment radius at lower masses and underestimates it at larger masses compared to our work. If the scaling laws presented in M10 are corrected there is better overall agreement to our work. However, for masses below \SI{0.46}{\Mearth} the iron mass fraction is larger than $1$ making the validity of these results at lower masses questionable. Generally, the radii inferred from our scaling laws agree with those presented in M10 within \SI{6.5}{\percent}. Given that future missions like PLATO \citep{Rauer2014} are expected to reduce uncertainty in the measured planetary radius below three to five percent these differences are relevant for validating the effect of GI on their composition. Another aspect when deriving a fragment mass and composition for given impact conditions that has received less attention so far is the total colliding mass involved. The bottom panel of Figure~\ref{fig:sl_radius_vs_frg_mass} shows the (pre-impact) mass of the target for a given fragment mass and radius. As expected for a given fragment mass the required target mass increases with increasing impact velocity. In the most extreme case, for an impact velocity of \SI{160}{\kilo\meter\per\second}, the target mass varies between \SI{40}{\Mearth} and \SI{93}{\Mearth}. 

Clearly, defining a minimum radius (or maximum iron mass fraction) of an exoplanet due to collisional stripping of its mantle based solely on the impact velocity as proposed in M10 is incomplete as it does not account for unrealistically large target and impactor masses that can be required to obtain very massive super-Mercuries. It is therefore desirable to account for the total colliding mass in addition to the impact velocity. Given the range in architectures of exoplanetary systems and their host stars, realistic impact conditions can vary greatly; a suitable criterion accounts for this diversity.

\subsection{Implications for iron-rich exoplanets}
\label{sec:results_exo_planets}

In Figure~\ref{fig:composition_vs_mass_frg} we present the composition of the iron-rich exoplanets presented in Table~\ref{tab_ExoplanetProperties} and the post-impact fragments obtained in our simulations. The exoplanets' compositions are inferred from interior modelling as described in Section~\ref{sec:methods_interior_models}. The grey lines represent the maximum iron mass fraction predicted by M10 in Figure 2b and by the correct equations as described in the Appendix~\ref{appendix:discussion_results_m10}. The shaded grey areas show the possible compositions predicted by our scaling laws for equal-mass collisions with $\SI{0.1}{\Mearth} \leq M_\textnormal{tar} \leq \SI{10}{\Mearth}$ and $\SI{20}{\kilo\meter\per\second} \leq v_\textnormal{imp} \leq \SI{100}{\kilo\meter\per\second}$ (dark grey) and $\SI{0.1}{\Mearth} \leq M_\textnormal{tar} \leq \SI{20}{\Mearth}$ and $\SI{20}{\kilo\meter\per\second} \leq v_\textnormal{imp} \leq \SI{160}{\kilo\meter\per\second}$ (light grey). The curves that envelope this region are the minimum and maximum iron mass fraction for these impact conditions. Because the minimum radius curves derived in M10 do not depend on the total colliding mass they over-predict the iron mass fractions for larger fragment masses compared to this work and give unrealistically high values. For small fragment masses on the other hand the iron mass fraction predicted by M10 is lower than in our work because the impact velocities in M10 are limited to \SI{80}{\kilo\meter\per\second}. We compare our results with observations in the M-R diagram presented in Figure~\ref{fig:radius_vs_frg_mass}. The measured masses and radii of the iron-rich exoplanets are taken from \citet{Otegi2020} and are listed in Table 1. Also presented are the radii of the fragments obtained in our simulations as well as the minimum and maximum radius of the fragments inferred from our scaling laws. The radii are inferred from the interiors models presented in section~\ref{sec:methods_interior_models}. The dashed curves show different compositions ranging from pure-rock to pure-iron planets. 

The inferred exoplanets' compositions fall within the range of the predicted compositions based on our impact simulations. Furthermore, our scaling laws show that five out of nine exoplanets have a composition that is consistent with mantle stripping involving a target mass of $\leq \SI{10}{\Mearth}$ and impact velocities below \SI{100}{\kilo\meter\per\second}. The other four agree with these impact conditions within the compositional uncertainly but likely require a target mass of $\leq \SI{20}{\Mearth}$ and impact velocities $\leq \SI{160}{\kilo\meter\per\second}$ to explain their larger masses and extreme compositions with GI. This suggests that in order to further constrain the impact conditions more accurate  observational data are required.

Since all of the iron-rich exoplanets considered in this work are orbiting very close to their host star and therefore have orbital velocities between \SI{130}{\kilo\meter\per\second} and \SI{260}{\kilo\meter\per\second} the limiting impact velocity of \SI{80}{\kilo\meter\per\second} assumed in M10 is too restrictive. Even velocities up to \SI{160}{\kilo\meter\per\second} seem plausible. It is possible that these metal-rich exoplanets formed farther out and migrated inwards. If these planets formed in resonant chains \citep{Izidoro2017} such an instability can lead to high-velocity encounters and even larger impact velocities (e.g., \citealt{Scora2020, Clement2021}). Therefore, exoplanets with extreme iron mass fractions could be a few lonely survivors, as has been suggested for Mercury \citep{Clement2021}. This hypothesis is consistent with the fact that the iron-rich exoplanets in our sample are single planets or correspond to the innermost planet in systems with a small number of planets with significant distancing between them. Two exceptions are the systems Kepler-80 and Kepler-107 that are rather compact and near resonance.

However, a target mass of $\sim \SI{20}{\Mearth}$ is more difficult to explain. While the formation of such massive super-Earths may not be excluded a priori, rather specific conditions are required. First, enough mass in the form of refractory elements is required at the location where such planets form. Second, once a body becomes so massive it is expected to accrete a substantial layer of hydrogen (H) and helium (He) from the proto-planetary disk and therefore have a composition similar to a giant planet rather than a super-Earth. Due to the close proximity to the host star and the corresponding high temperatures and stellar irradiation a fraction of this envelope could have been lost over time (e.g., \citealt{Owen2019}). Indeed, in some cases we find that a massive iron-rich body has a sub-Neptune companion. If still present at the time of the collision, a primordial H-He envelope can also be lost during the mantle stripping GI (e.g., \citealt{Kegerreis2020}). However, the presence of a massive envelope would increase the planetary mass and therefore more material must be stripped in a given collision. As a consequence, higher impactor masses and impact velocities are required to obtain a desired composition. 

While the close proximity of the iron-rich exoplanets in our sample to their host stars promote disruptive collisions, the corresponding short orbital periods imply that part of the ejected debris could be re-accreted rapidly. For Mercury, \citet{Benz2007} suggested that the ejecta could be in the form of small droplets that can be swept away via Poynting-Robertson drag \citep{Benz2007} or Solar winds \citep{Spalding2020}. Furthermore, since the observed exoplanets orbit close to their stars, their Hill radii are very small and tidal forces could play an important role in preventing re-accretion of ejecta.

Finally, the initial iron mass fraction (pre-impact) can vary, e.g., if the host star is enriched in iron as proposed in \citet{Adibekyan2021}. For impact energies below the transition to rapid mass loss the bound mass is not affected by the initial iron mass fraction (see Section \ref{sec:results_sim} for details). Likewise, the post-impact composition in this regime shows a similar behaviour as the initial iron mass fraction enters Equation~\ref{eqn:iron_mass_frac_vs_q_imp} as a constant. Generally, for a higher initial iron mass fraction less material has to be removed in the collision. Therefore, a lower total colliding mass and impact velocity are required to obtain a desired iron mass fraction which facilitates the formation of very massive super-Mercuries. Since the transition to rapid mass loss shows some dependency of the initial iron mass fraction, it would be desirable to investigate a range of different core mass fractions in future studies.

\section{Summary and conclusions}
\label{sec:summary}
In this paper we present a large suite of SPH simulations of impacts between super-Earths in order to investigate the efficiency of mantle stripping in GI. Motivated by the observation of super-Mercuries with masses $> \SI{5}{\Mearth}$ we study head-on impacts between super-Earths with masses between \SI{1}{\Mearth} and \SI{19.04}{\Mearth} and impact velocities between two to six times the mutual escape velocity. From these simulations we determine the bound mass, which corresponds to the post-impact planetary mass, as well as the iron mass fraction and radius. For each combination of target and impactor mass we also calculate the critical specific impact energy for catastrophic disruption $Q_{\textnormal{RD}}^{*}$. We then propose new scaling laws for the post-impact mass and composition as well as \QRD that more accurately represent our data than prior work that investigated lower mass bodies (SL09 and M09). Generally we find that at low specific impact energies more mass remains bound and the iron mass fraction therefore remains lower than predicted in M09. In the case of collisions occurring at specific impact energies above \QRD more mass is stripped and correspondingly the iron mass fraction is higher than expected from M09.

We then combine these scaling laws to obtain equations that relate the impact conditions, e.g., target mass, impactor-to-target mass ratio and impact velocity, to the post-impact mass and composition of a planet. These equations allow one to predict collision outcomes for different parameters without the need of performing a large suite of impact simulations. We find that obtaining a desired fragment mass and composition requires higher impact velocities than predicted in M10. Furthermore, for a given impact velocity we predict higher iron mass fractions for low fragment masses and lower iron mass fractions for high fragment masses compared to M10. This is reflected in the minimum radius of a super-Earth where we observe differences to M10 of up to \SI{6.5}{\percent}. Additionally, we find that M10 predicts an iron mass fraction larger than $1$ for smaller masses because the scaling law for the iron mass fraction presented in M09 does not correctly extrapolate to $1$ for very large impact energies. Considering that future missions are expected to reduce the uncertainty in observed planetary radius to \SIrange{3}{5}{\percent} these differences are relevant when interpreting the origin of iron-rich planets. Finally, we find several errors in the equations of M10 (see Appendix~\ref{appendix:discussion_results_m10} for details) that can result in differences of eight percent in the predicted planetary radius, once corrected, and therefore are comparable to the uncertainty of current observations. Future work should use our new results, or at the very least, include the corrections to M10 presented in Appendix~\ref{appendix:discussion_results_m10}.

In order to connect our findings to observations we compare the post-impact bodies found in our simulations to a sample of iron-rich exoplanets and find that they are in good agreement. A giant impact occurring in these exoplanets' early history therefore could indeed explain their large iron mass fractions. From the scaling laws derived in this work we also mark the outcome of collisions for different ranges of pre-impact mass and impact velocity by deriving a minimum and maximum radius and iron mass fraction respectively. This allows to further constrain possible impact conditions required to obtain these exoplanets' compositions. We find that five out of nine exoplanets in our sample are consistent with impacts involving a $<\SI{10}{\Mearth}$ target and impact velocities below \SI{100}{\kilo\meter\per\second}. The other four are constrained by target masses $< \SI{20}{\Mearth}$ and impact velocities below \SI{160}{\kilo\meter\per\second}.

Finally, we propose the minimum and maximum radius for a rocky exoplanet of a given mass due to collisional stripping of its mantle that accounts for both the required impact velocities and total colliding mass. Between the two curves are all possible compositions and radii of exoplanets due to mantle stripping GI for given constraints on the impact conditions. Due to the diverse architecture of exoplanetary systems and their host stars constraints on the impact conditions can vary substantially. Therefore, we provide \textsc{python} scripts (see data availability statement) that allow users to choose their desired minimum and maximum impact velocity and total colliding mass respectively, allowing one to constrain possible impact conditions for a broad variety of observed exoplanetary systems.\\

Our main conclusions can be summarised as follows:
\begin{itemize}
    \item For collisions in the super-critical disruption regime ($ Q_{\textnormal{R}} > 1.26~Q_{\textnormal{RD}}^{*}$) details of the EOS, in particular the treatment of the liquid phase, are crucial for investigating mantle stripping GI.
    \item Prior work has overestimated the mantle stripping of large rocky exoplanets at lower specific impact energies, while underestimating this at higher energies.
    \item Impacts occurring at specific impact energies below 0.5~\QRD have little effect on a planet's mass and iron mass fraction.
    \item The formation of very massive super-Mercuries requires both high impact velocities as well as very massive pre-impact bodies. The existence of such planets implies very specific formation conditions, and hence we expect such planets to remain rare.
    \item The observed masses and inferred compositions of super-Mercuries are consistent with our simulation results implying that GI play an important role during their formation.
    \item The new M-R curves (for the software see data availability statement) allow one to bracket the initial mass and velocity allowed for the GI formation of rocky exoplanets.
    \item Break-up of resonant chains may result in architectures favourable for the formation of super-Mercury exoplanetary systems (close proximity to host star, high relative velocities and large masses). 
\end{itemize}

Our understanding of super-Mercuries is still incomplete and is limited by the small number of observed metal-rich exoplanets and the data accuracy. Ongoing and future space missions like TESS and PLATO will substantially reduce the uncertainty of observations on mass and radius which gives tighter constraints on the composition of observed exoplanets. This will allow to better distinguish super-Mercuries from super-Earths and to assess the occurrence rate of such planets in the galaxy. In combination with more accurate data regarding the stellar properties and architecture of systems hosting super-Mercuries this will tighten the constraints on possible formation scenario and improve our understanding of the role of giant impacts.

\section*{Acknowledgements}

We thank John Chambers for valuable suggestions and comments that helped to substantially improve the paper. We also thank Caroline Dorn for helpful discussions.
This work has been carried out within the framework of the National Centre of Competence in Research PlanetS supported by the Swiss National Science Foundation. The authors acknowledge the financial support of the SNSF.
We acknowledge access to Piz Daint or Eiger@Alps at the Swiss National Supercomputing Centre, Switzerland under the University of Zurich's share with the project ID UZH4.
This publication makes use of The Data \& Analysis Center for Exoplanets (DACE), which is a facility based at the University of Geneva (CH) dedicated to extrasolar planets data visualisation, exchange and analysis. DACE is a platform of the Swiss National Centre of Competence in Research (NCCR) PlanetS, federating the Swiss expertise in Exoplanet research. The DACE platform is available at \url{https://dace.unige.ch}.

\section*{Data Availability}
The data underlying this article will be shared on reasonable request to the corresponding author. The scaling laws based on this data are implemented in \textsc{python} scripts that are available at Zenodo \url{https://doi.org/10.5281/zenodo.6592438}.

\section*{Software}
In the present work we used the following software: \textsc{Gasoline} \citep{Wadsley2004}, \textsc{ballic} \citep{Reinhardt2017}, \textsc{EOSlib} \citep{Meier2021a, Meier2021b}, \textsc{M-ANEOS} \citep{Thompson2019, Stewart2019, Stewart2020}, \textsc{skid}\footnote{\url{https://github.com/N-BodyShop/skid}}, \textsc{numpy} \citep{Harris2020}, \textsc{scipy} \citep{Virtanen2020}, \textsc{matplotlib} \citep{Hunter2007}

\bibliographystyle{mnras}
\bibliography{main} 


\appendix

\section{Discussion of the results from M10}
\label{appendix:discussion_results_m10}

\begin{figure}
    \includegraphics[width=\columnwidth]{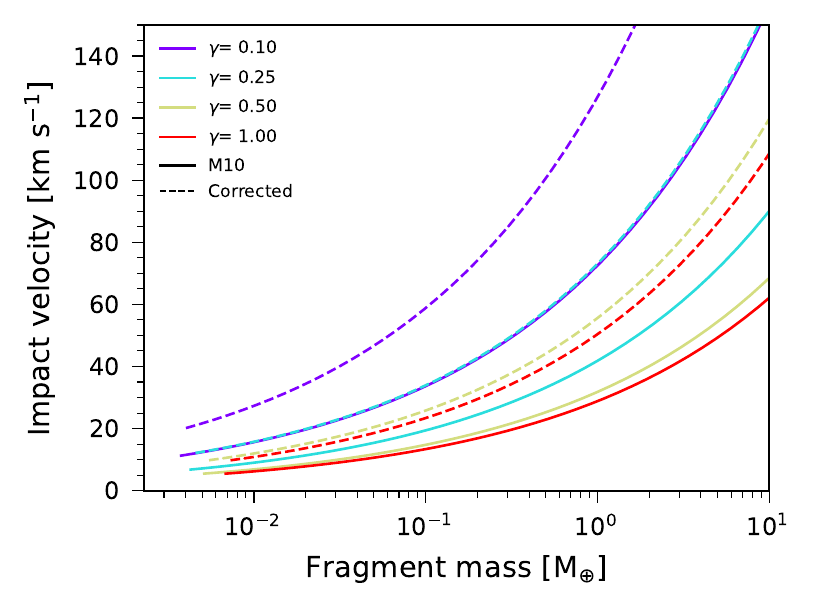}
    \includegraphics[width=\columnwidth]{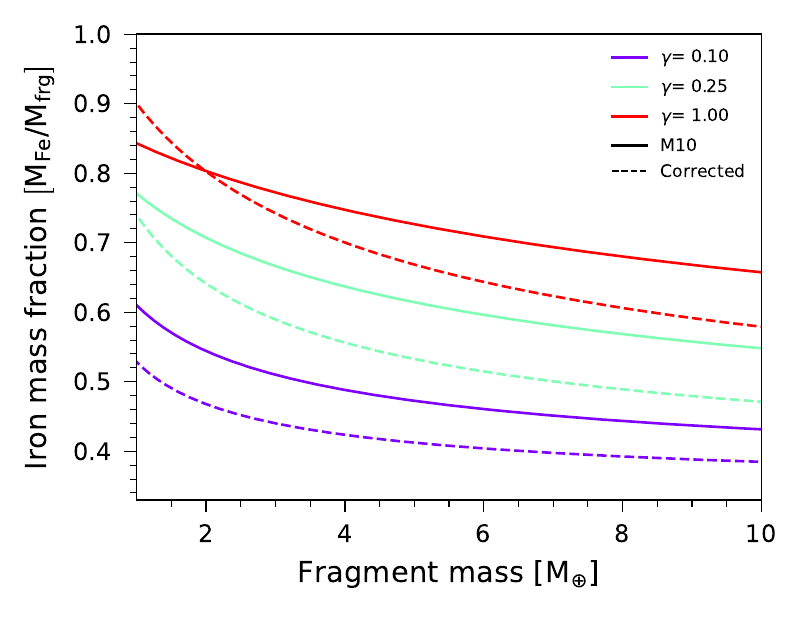}
    \caption{\textbf{The effect of the errors in the equations of M10 on their results.} The solid lines are the curves shown in Figure 1b and 2b of M10. The dashed lines are the results obtained from the corrected equations derived from the original scaling laws in M09. The top panel shows the impact velocity required to obtain a desired fragment mass with an iron mass fraction of $0.7$ for different impactor-to-target mass ratios $\gamma$ (purple: $\gamma=0.1$, cyan: $\gamma=0.25$, yellow: $\gamma=0.5$ and red: $1.0$). The mistake in the prefactor (see Appendix~\ref{appendix:discussion_results_m10} for details) results in a severe underestimation of the impact velocity required to produce a Mercury-like planet via collisional stripping of the mantle. The bottom panel shows the iron mass fraction of the largest fragment for an impact velocity of \SI{80}{\kilo\meter\per\second} and different impactor-to-target mass ratios (purple: $\gamma=0.1$, light green: $\gamma=0.25$ and red: $\gamma=1.0$). Compared to the curve shown in Figure 2b of M10 the corrected equations predict a lower iron mass fraction except for equal mass collisions where it is higher for small fragment masses.}
    \label{fig:plot_marcus10}
\end{figure}

In M10 equations for the impact velocity required to produce a given fragment mass and composition due to collisional stripping are derived by combining the scaling laws presented in SL09 and M09. Trying to compare their results to our work we find several typos and calculation errors that make it difficult to reproduce the figures 1b and 2b in M10. First, we find that the prefactor of $10.5$ in Equation 1 of M10 is wrong and should be $18.75$ because they did not account for a factor $0.5$ that enters the equation from the kinetic energy term. The impact velocities in Figure 1 of their paper are therefore underestimated by approximately a factor of two. Second, the curves in Figure 2 of the paper can only be reproduced if the exponent in Equation 2 (in M10) is $0.2606$ rather than $0.606$ (or $1/1.65$) as written in the paper. It seems that M10 made a simple typo when calculating these curves for the plots. Finally, many prefactors and exponents require higher precision than given in the paper. Most importantly, the prefactor $-1.2$ in Equation 2 should be $-1.1584$ in order to accurately reproduce the curves in M10.

These mistakes have a profound impact on the results of M10. In Figure~\ref{fig:plot_marcus10} we compare the curves obtained using the correct prefactors and exponents to the ones shown in Figure 1b and 2b of M10. As mentioned above the impact velocity required to obtain a desired fragment mass with an iron mass fraction of Z$_\textnormal{Fe} = 0.7$ via collisional stripping of the planet's mantle is about twice as large as suggested in Figure 1b of M10. Likewise the iron mass fraction of the largest fragment resulting from a collision at an impact velocity of \SI{80}{\kilo\meter\per\second} is very different once the correct equations are used. For example, in case of an equal mass collision the iron mass fraction is $0.9$ rather than $0.84$ for a fragment mass of \SI{1}{\Mearth} and $0.66$ rather than $0.58$ for a fragment mass of \SI{10}{\Mearth}. This in turn affects the fragment radius as well as the total colliding mass required to obtain a given composition by collisional stripping (see Figure~\ref{fig:sl_radius_vs_frg_mass}). The difference in radius between the correct and the published results of M10 are $\leq \SI{8.5}{\percent}$ which is comparable to the uncertainty of current observations. Furthermore, the total colliding mass predicted by the published results is massively overestimated compared to the corrected equations as well as our results. Given the popularity and importance of the minimum radius of super-Earth presented in M10 in interpreting observational data (here some of the most cited papers: \citealt{Holman2010, Batalha2011, Lopez2013, Lopez2014, Rogers2015}) we suggest taking these corrections into consideration in future work.


\bsp	
\label{lastpage}
\end{document}